\pdfoutput=1
\documentclass[twocolumn,floatfix,preprintnumbers,citeautoscript,prl,superscriptaddress,amsmath,amssymb]{revtex4}

\usepackage{hyperref}
\usepackage{graphicx}
\usepackage{color}
\usepackage{amsmath}
\usepackage{amssymb}
\usepackage[utf8]{inputenc}

\def\be#1\ee{\begin{align}\begin{split}#1\end{split}\end{align}}
\def\beq#1\eeq{\begin{align}\begin{split}#1\end{split}\end{align}}

\newcommand\SU{\mathrm{SU}}
\newcommand\U{\mathrm{U}}

\newcommand{\Mp}{M_{\rm Pl}}

\begin{document}

\title{Quintessence Axion Revisited in Light of Swampland Conjectures}

\author{Masahiro Ibe}
\affiliation{Institute for Cosmic Ray Research, University of Tokyo, Kashiwa, Chiba 277-8582, Japan}
\affiliation{Kavli Institute for the Physics and Mathematics of the Universe (WPI), University of Tokyo, Kashiwa, Chiba 277-8583, Japan}

\author{Masahito Yamazaki}
\affiliation{Kavli Institute for the Physics and Mathematics of the Universe (WPI), University of Tokyo, Kashiwa, Chiba 277-8583, Japan}

\author{Tsutomu T.~Yanagida}
\affiliation{Kavli Institute for the Physics and Mathematics of the Universe (WPI), University of Tokyo, Kashiwa, Chiba 277-8583, Japan}
\affiliation{T.~D.~Lee Institute and School of Physics and Astronomy, Shanghai Jiao Tong University, Shanghai 200240, China}

\date{November, 2018}

\preprint{IPMU-18-0183}

\begin{abstract}
We point out that the swampland conjectures,  forbidding the presence of
global symmetries and (meta-)stable de Sitter vacua within quantum gravity,
pick up  a dynamical axion for the electroweak SU(2) gauge theory as
a natural candidate for the quintessence field.
The potential energy of the electroweak axion provides an attractive candidate for the dark energy. 
We discuss constraints from the weak gravity conjecture, from the conjecture of no global symmetry, 
and from observations, which can be satisfied elegantly in a supersymmetric extension of the standard model.
\end{abstract}

\maketitle

\bigskip\noindent
{\bf Introduction}

It has long been a challenging problem to identify any possible consequences of UV quantum gravity for IR physics. For this purpose there has been several conjectures (the so-called swampland conjectures \cite{Vafa:2005ui,Ooguri:2006in,Brennan:2017rbf}) in the literature, which claim necessary conditions for a low-energy effective field theory to have a consistent UV completion within theories of quantum gravity.

One of the most striking conjectures states that a stable de Sitter (dS) vacuum is forbidden in quantum gravity (see \cite{Dvali:2014gua,Dvali:2017eba,Sethi:2017phn,Danielsson:2018ztv,Obied:2018sgi} for recent discussion), as is articulated in the recent de Sitter swampland conjecture \cite{Obied:2018sgi} and its refinements \cite{Andriot:2018wzk,Garg:2018reu,Murayama:2018lie,Ooguri:2018wrx,Garg:2018zdg}.
The conjectures of \cite{Obied:2018sgi,Ooguri:2018wrx} are motivated by the no-go result of \cite{Maldacena:2000mw}, and is better motivated by the swampland distance conjecture 
\cite{Ooguri:2006in,Klaewer:2016kiy} in asymptotic regions of the parameter space where we have parametric control \cite{Dine:1985he,Ooguri:2018wrx,Hebecker:2018vxz}. While this dS conjecture is still speculative, the conjecture has triggered active discussions
in the community. It was further conjectured that there exists no non-supersymmetric anti-de Sitter (AdS) vacua \cite{Ooguri:2016pdq}.

 In this Letter, we first point out the inconsistency of the global symmetry with the quantum gravity makes the 
$\theta$-angle of the electroweak $\SU(2)$ gauge theory a physical parameter.
We then discuss the possibility that the electroweak axion can be regarded as the quintessence field generating the dark energy
which is essential to render the observed dark energy consistent with the conjectural absence of the dS vacua
\footnote{A similar discussion for the QCD axion \cite{Peccei:1977ur,Weinberg:1977ma,Wilczek:1977pj}
can be found in the recent paper \cite{Dvali:2018dce}.}.
We find impressive consistency with theoretical and observational constraints, 
if we consider a supersymmetric extension of the standard model,
where a phenomena, the Supersymmetric Miracle, plays crucial roles.

\bigskip\noindent
{\bf Electroweak Axion from Swampland Conjectures}

Let us first present our argument for the electroweak axion. We emphasize again that 
our argument is based on a few swampland conjectures.

In the electroweak $\SU(2)$ gauge theory, we consider the 
$\theta$-angle and the associated $\theta$-vacuum \cite{Callan:1976je} 
for each fixed angle $\theta$. One might argue that the $\theta$-angle in $\SU(2)$ gauge theory can be rotated away by anomalies for the global $B+L$ symmetry (here $B$ and $L$ are baryon and lepton numbers, respectively),
and hence is unphysical. However, an exact global symmetry is forbidden in theories of quantum gravity \cite{Misner:1957mt,Polchinski:2003bq,Banks:2010zn,Harlow:2018tng},
and we expect that the $B+L$ symmetry is broken by higher-dimensional operators in the standard model,
such as the dimension $6$ operator $qqql$ (where $q$ and $l$ are the quarks and the leptons, respectively).
In the presence of $B+L$ breaking, the $\theta$-angle of the electroweak $\SU(2)$  gauge theory can no longer be rotated away.

The $\theta$-vacuum is a stable vacuum, and we can consider such vacuum for any value of $\theta$ as a boundary condition of the universe. Here, we assume that the UV quantum gravity provides a Lagrangian of the effective 
low energy quantum field theory while it does not control the boundary condition of the low energy theories.
 Thus, the vacuum energy is generically of order of the dynamical scale of the electroweak theory. 
The existence of such vacua contradicts swampland conjectures, that is, if the vacuum energy is positive this violates the refined swampland dS conjecture of \cite{Ooguri:2018wrx}.
The inconsistency with the swampland dS conjecture often invokes the quintessence field to tilt the scalar potential at the vacuum slightly.

The above arguments point us an interesting possibility of identifying
the quintessence field with a dynamical axion for the electroweak $\SU(2)$ gauge theory.
Namely the $\theta$-angle is promoted to the electroweak axion $a$,
which couples to the electroweak gauge field as:
\be
\mathcal{L}_{aF\tilde{F}}=\frac{a}{32 \pi^2 f_a}\textrm{Tr} F_{\mu\nu} \tilde{F}^{\mu\nu},
\label{L_aFF}
\ee
where $f_a$ is the decay constant.
Here, it should be emphasized that we make a very strong assumption that 
 the electroweak axion does not have any other interaction terms beyond the one 
in Eq.~\eqref{L_aFF}. If there are other shift symmetry breaking terms, the axion potential 
may have many local minima with positive vacuum energy in general, which are also excluded 
by the swampland conjectures.
Besides, the additional contributions to the scalar potential of the electroweak axion can spoil the
numerical success of the size of the dark energy generated by the electroweak instanton 
which we will see shortly.

Incidentally, the presence of two light axions (the electroweak axion and the QCD axion)
has unexpected virtue, in that the two light axions removes the tension between Majorana neutrinos and the conjectured absence of the non-supersymmetric stable 
AdS vacua, see section 5.2.2 of \cite{Ibanez:2017kvh}.

\bigskip\noindent
{\bf Electroweak Axion as Quintessence}

The shift symmetry $a\to a+\textrm{(const.)}$ of the axion is broken in the presence of the $qqql$ operator,
and non-perturbative instanton effects generate an axion potential \cite{tHooft:1976snw}, 
\be
V(a)=\Lambda_a^4 \left(1-\cos\frac{a}{f_a} \right), \label{V_a}
\ee
in the one-instanton approximation.
Here $\Lambda_a$ is the dynamically-generated scale,
whose naive estimate is given by
\be
\Lambda_a^4&=\Mp^4 \, e^{-\frac{2\pi}{\alpha_2(\Mp)}} \simeq 10^{-130} \Mp^4  \ll \Mp^4 ,
\label{Lambda_a}
\ee
where we use the value of the electroweak coupling constant $\alpha_2=g_2^2/(4 \pi)$ at the Planck scale $\alpha_2(\Mp)=1/48$,
and $\Mp \simeq 2\times 10^{18} \textrm{GeV}$ is the reduced Planck scale.
The axion potential is dominated by small-size instanton contributions since it requires 
insertions of the higher-dimensional $qqql$ operators.

While the potential \eqref{V_a} is problematic \cite{Murayama:2018lie} in the original version \cite{Obied:2018sgi} of the de Sitter swampland conjecture (see also \cite{Denef:2018etk,Conlon:2018eyr,Choi:2018rze,Hamaguchi:2018vtv}), 
the refinement \cite{Ooguri:2018wrx} is compatible 
with the axion potential as it is vanishing at the minimum at $a = 0$ (mod $2\pi f_a$)%
\footnote{The axion potential may take a negative value at the minimum which can be consistent with the swampland conjectures 
on stable AdS vacua as long as we assume that they are unstable quantum mechanically.}.
In this Letter we assume the refined dS conjecture of \cite{Ooguri:2018wrx}.

The curious observation is that the
value of $\Lambda_a$ is close to the current energy scale of the 
dark energy \cite{Nomura:2000yk,McLerran:2012mm}, 
\be
\Lambda_0^4\simeq 10^{-120} \Mp^4 \simeq O(\textrm{meV})^4.
\label{Lambda_0}
\ee
This estimate shows that the identification of the idea of the \emph{quintessence axion} \cite{Fukugita:1994hq,Frieman:1995pm,Choi:1999xn}
is highly successful, where the dark energy can be explained by the dynamical electroweak axion. 
Here the shift symmetry of the axion ensures the flatness of the quintessence potential
against possible quantum corrections.

While the conjectural absence of the de Sitter vacua 
has already been used as motivations for quintessence  \cite{Ratra:1987rm,Wetterich:1987fm,Zlatev:1998tr}
explanation of dark energy (see \cite{Obied:2018sgi,Heisenberg:2018yae,Heisenberg:2018rdu,DAmico:2018mnx,Agrawal:2018rcg} for recent discussion), we have shown 
here that the conjecture brings a natural candidate for the quintessence, the \emph{quintessence axion}. 

\bigskip\noindent
{\bf Constraints from the Weak Gravity Conjecture}

To this point, we have motivated the electroweak quintessence axion \cite{Nomura:2000yk,McLerran:2012mm}
from the one of the swampland conjectures, namely the absence of the dS and AdS vacua. It turns out, however, this scenario is some tension with another swampland conjecture,
the weak gravity conjecture \cite{ArkaniHamed:2006dz}.

The weak gravity conjecture sets an upper bound on the decay constant
\be
f_a\lesssim \frac{\Mp}{S_{\rm inst}} ,
\label{WGC_orig}
\ee
where $S_{\rm inst}\simeq 2\pi / \alpha_2(\Mp)$ is the size of instanton action. 
Notice we have chosen the value of $\alpha_2$ here to be evaluated at the UV scale
and not at the IR scale, since as we have described above the 
$\theta$-angle is physical only after taking into account 
higher-dimensional $(B+L)$-breaking operators.

Since
$S_{\rm inst}\sim O(100)$ for the electroweak instanton, we obtain the bound
\be
f_a\lesssim O(10^{16}\, \textrm{GeV}).
\label{WGC}
\ee
This is parametrically smaller than the value $f_a\sim \Mp \sim 10^{18}\, \textrm{GeV}$,
which was assumed in \cite{Nomura:2000yk}.

While the energy scale of dark energy \eqref{Lambda_a} is not affected by the value of the
decay constant $f_a$, such a low value of $f_a$ predicts a larger value for the mass $m_a$ for
the quintessence axion \cite{Nomura:2000yk}:
\be
m_a^2&\simeq \frac{\Lambda^4}{f_a^2} \simeq \frac{H_0^2 \Mp^2}{f_a^2}
\gtrsim H_0^2 =(2\times 10^{-33} \textrm{eV})^2,
 \label{m_a}
\ee
where $H_0$ is the present-day value of the Hubble constant.
This means that the quintessence field has already started rolling down the potential,
and the slow-roll condition  $\Mp |V'(a)| \ll V(a)$ is no longer satisfied.
This is problematic for the axion as a dark energy candidate.

One possibility to circumnavigate this problem is to fine-tune the initial condition of the 
quintessence to be close to  the local maximum $a\sim \pi f$, so that the quintessence,
while it has already started rolling down the potential, is still located close to the local maximum. 

This hilltop quintessence scenario \cite{Dutta:2008qn}, however, has a severe
fine-tuning problem in our context. 
Indeed, if we start with a small initial value of the displacement $\delta a=a-\pi f_a$, we expect that $\delta a$
will grow.
We can estimate this growth from the equation of motion for the axion
\be
\ddot{a}+ 3H_0 \dot{a}= -V'(a), \label{a_eom}
\ee
where $H_0$ is the Hubble constant. Here we
use the present-day value of the Hubble constant since we are interested 
in the era when the dark energy dominates the potential energy of the Universe.
When the displacement $\delta a$ is small, we can approximate Eq.~\eqref{a_eom} as
\be
\ddot{\delta a}+3H_0\dot{\delta a}= \frac{3 H_0^2 \Mp^2}{f^2} \delta a,
\ee
where we used $\Lambda^4\simeq 3 H_0^2 \Mp^2$.
Solving this equation we estimate that $\delta a$ grows exponentially 
($\delta a(t) \propto \exp(\sqrt{3} H_0 M_{\rm Pl} t /f)$),  
so that after the axion starts rolling the total growth until today ($t\sim O(H_0^{-1})$) 
is of order $\exp(O(\Mp/ f))=\exp(O(100))$.
In order to ensure slow-roll condition until now,
initial displacement when the axion starts rolling needs to be constrained in a
tiny window \footnote{See also \cite{Choi:1999xn}. Similar exponential fine-tuning was discussed recently in \cite{Fukuda:2018haz} for curvatons.}
\be
\frac{|\delta a|}{f_a} \ll \frac{f_a}{\Mp}  e^{-O(100)}.
\label{a_window}
\ee
One moreover expects that such an extreme fine-tuning is incompatible with the 
fluctuations generated during the inflationary era.

We therefore conclude that the hilltop quintessence scenario,
as required by the small value $f_a\lesssim O(10^{16} \, \textrm{GeV)}$ of the decay constant \eqref{WGC}, is strongly disfavored.

\bigskip\noindent
{\bf Inclusion of Heavy Particles}

Despite the difficulties mentioned above, 
we can save the electroweak quintessence axion scenario
by taking into account the effects of heavy particles.

Recall that in the weak gravity conjecture bound \eqref{WGC_orig}
we have substituted the value $S_{\rm inst}\sim O(100)$
which was obtained by renormalization group (RG) running in 
the $\SU(2)$ gauge group all the way up to the Planck scale,
assuming that there are  no massive particles modifying the RG running of $\alpha_2$.
However, there is no strong justification for this assumption.

Suppose that the RG running of $\alpha_2$ is changed so that 
we have a larger value of $S_{\rm inst}$ at the Planck scale.
For example, if $S_{\rm inst}=O(1)$ we have
$f\sim \Mp$.
The axion mass \eqref{m_a} can then be taken to be $m_a\sim H_0$, 
which makes the fine-tuning condition on the field value in Eq.\,(\ref{a_window})
much less severe to achieve  the slow-roll conditions around the hill-top region
\footnote{For $f_{a} = {\cal O(M_{\rm Pl})}$, the axion potential can be modulated by the
higher harmonics or gravitational instanton effects.
Since the domain of the axion field is compact, the axion potential have local maximum 
even with those effects, and hence, the slow-roll conditions can be satisfied around 
the hill-top region without severe fine-tuning as long as the overall scale of the 
axion potential is not altered significantly.}.
It should be also noted that the value $f_a\lesssim \Mp$ of the decay constant is also compatible with the refined swampland distance conjecture \cite{Ooguri:2006in,Klaewer:2016kiy}, 
which restricts the field range for the axion to be $\Delta a \lesssim O(\Mp)$.

There is, however, still one problem if one changes  
the value of the coupling constant at the Planck scale $\alpha_2(\Mp)$---the same 
coupling constant appears in the dynamical energy scale $\Lambda_a$
of the quintessence \eqref{Lambda_a}, and once we modify the value of 
$\alpha_2(\Mp)$ we are spoiling the crucial observation $\Lambda_a \simeq \Lambda_0$,
which was the very starting point for our quintessence axion scenario%
\footnote{The deviation from $\Lambda_a \simeq \Lambda_0$ might 
open up another interesting application of the EW axion 
such as a candidate for the ultralight (fuzzy) matter~\cite{Du:2016zcv}.
}.

Interestingly, this problem is solved elegantly in supersymmetric theories,
which we turn next.

\bigskip\noindent
{\bf Supersymmetric Miracle}

Let us consider the minimal supersymmetric standard model
(MSSM). The dynamical scale of the axion potential is computed in the instanton calculus as \cite{Nomura:2000yk}
\be
\Lambda_a^4
&\simeq e^{-\frac{2\pi}{\alpha_2(\Mp)|_{\rm MSSM} }} \epsilon^{10} m_{\rm SUSY}^3 \Mp  .\label{Lambda_a_SUSY}
\ee
Here $\alpha_2(\Mp)|_{\rm MSSM}$ is the $\SU(2)$ gauge coupling constant at the Planck scale for the MSSM.
The mass scale $m_{\rm SUSY}\simeq O(\textrm{TeV})$ is the scale for spontaneous supersymmetry breaking,
whose exact value is not too important for what follows.

In the MSSM, the $B+L$ symmetry is broken by Planck-suppressed dimension $5$ operators $QQQL$  \cite{Sakai:1981pk,Weinberg:1981wj}.
These operators induce too rapid proton decay for $m_{\rm SUSY}\simeq O(\textrm{TeV})$.
To suppress the dimension $5$ operators, we assume the Froggatt-Nielsen (FN) symmetry
\cite{Froggatt:1978nt}, following  \cite{Nomura:2000yk}.
For now, we assume the FN symmetry is a global symmetry \footnote{Another possibility is to gauge a discrete subgroup of the FN symmetry. For example, when the FN charges are chose as in Table 2 of \cite{Nomura:2000yk} the $Z_{10}$ subgroup is anomaly free and can be gauged.}. 
We also assume the global $R$-symmetry given in  \cite{Nomura:2000yk}.
The FN symmetry is spontaneously broken by $\epsilon$, where the value of $\epsilon$ is taken to be 
$\epsilon \simeq 1/17$ \cite{Buchmuller:1998zf} to explain the quark/lepton masses and mixing angles.
With the charge assignment in  Table 2 of \cite{Nomura:2000yk}, we confirm that the 
proton decay via the $QQQL$ operators are sufficiently suppressed.

If do we not include any heavy particles beyond the MSSM we can substitute the value $\alpha_2(\Mp)|_{\rm MSSM}\simeq 1/23$, so that we obtain
\be
\Lambda_a^4
&\simeq \left(\frac{\epsilon}{1/17}\right)^{10} (\textrm{meV})^4 ,\label{cc_current}
\ee
which nicely reproduces the scale $\Lambda_0$ of the dark energy \cite{Nomura:2000yk}.

Suppose that  we include a pair of heavy massive particle $X, \bar{X}$ 
in some representation $R$ of $\SU(2)$ gauge group.
When the mass of $X, \bar{X}$ is at an intermediate energy scale $M_X$,
then the RG running of the coupling constant in the one-loop approximation is modified as
\be
\alpha_2^{-1} (\Mp)|_{X\bar{X}}=\alpha_2^{-1} (\Mp)\big|_{\rm MSSM} + \frac{2 T_R}{2 \pi} \log \frac{M_X}{\Mp},
\ee
where $T_R$ is the Dynkin index of the representation $R$.
Such heavy particles also change the 
zero modes, so that we need to insert operators $M_X X \bar{X}$ to
cancel the instanton zero modes, leading to the factor
$(M_X/\Mp)^{2T_R}$. Interestingly, these two effects cancel with each other for a supersymmetric theory,
leaving the dynamical scale $\Lambda_a$ invariant:
\be
\Lambda_a^4|_{X\bar{X}}
&\simeq e^{-\frac{2\pi}{\alpha_2(\Mp)|_{X\bar{X}} }} 
\left(\frac{M_X}{\Mp}\right)^{2T_R} \epsilon^{10} m_{\rm SUSY}^3 \Mp \\
&= c\, e^{-\frac{2\pi}{\alpha_2(\Mp)|_{\rm MSSM } }}  \epsilon^{10} m_{\rm SUSY}^3 \Mp \simeq \Lambda_a\big|_{\rm MSSM}^4.\label{Lambda_preserved}
\ee
We call this cancellation ``Supersymmetric Miracle".
Due to the Supersymmetric Miracle, $\alpha_2(M_{\rm Pl})$ 
can be achieved while the relation $\Lambda_a\simeq \Lambda_0$ intact.
While the one-instanton approximation is not reliable for a large value of the coupling constant $\alpha_2(\Mp)\sim 2\pi$,
we expect that the estimation of the energy scale $\Lambda_a$ \eqref{cc_current} will not be 
significantly affected by this subtlety.

Once $\alpha_2(M_{\rm Pl})\sim 2\pi$ is achieved, the size of the instanton action is of $S_{\rm inst} = O(1)$,
which makes the large decay constant, $f_a \sim M_{\rm Pl}$, marginally consistent with the weak gravity conjecture in 
Eq.\,\eqref{WGC_orig}.
Consequently, the axion mass $m_a$
\be
m_a^2&\simeq \frac{\Lambda^4}{f_a^2} \\
&\simeq \left(\frac{\epsilon}{1/17}\right)^{10} 
 \left(\frac{2\times 10^{18}\textrm{ GeV}} {f_a} \right)^2 (7\times 10^{-34} \textrm{eV})^2,
 \label{m_current}
\ee
is achieved.
In this way, the electroweak action serves as the slow-rolling quintessence field
consistently with the weak gravity conjecture.

\begin{figure}[t]
\includegraphics[trim=0 0 0 0,scale=0.6]{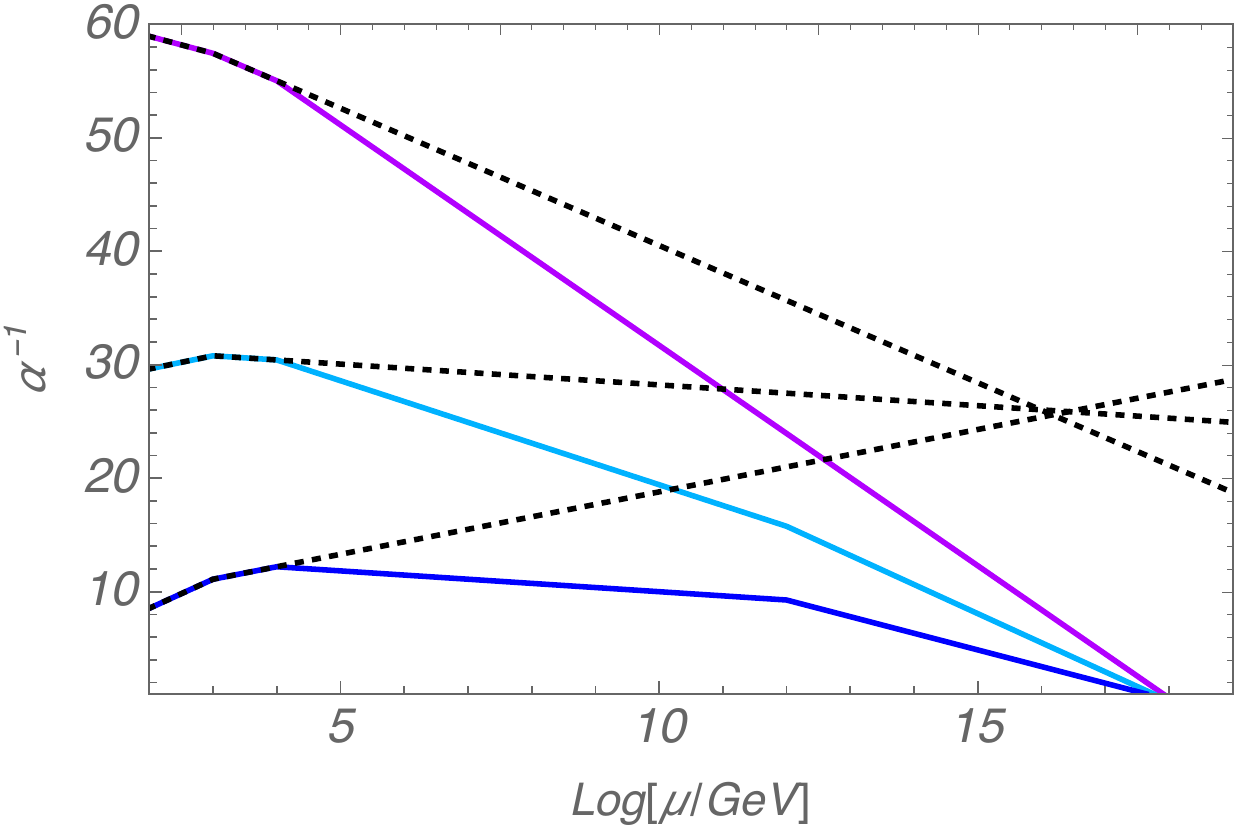}
\caption{The one-loop RG running of the $\SU(3)\times \SU(2)\times \U(1)$ gauge coupling constants as a function of the energy scale $\mu$.
We color $\alpha_1^{-1}, \alpha_2^{-1}, \alpha_3^{-1}$ by purple, light blue, dark blue, respectively.
We can achieve both (1) $\alpha_2(\Mp)\simeq 2\pi$ and
(2) gauge coupling unification at the Planck scale $\Mp$ by
including an $\SU(2)$ triplet and an $\SU(3)$ octet at $10^{12} \textrm{GeV}$,
together with $4$ pairs of GUT-like multiplets $\bf{5}$ and $\bar{\bf{5}}$ at $\sim 10\, \textrm{TeV}$.
The RG running in this case, with the supersymmetry breaking scale $m_{\rm SUSY}\simeq 1\, \textrm{TeV}$,
is shown as solid lines. This is contrasted with the case of the MSSM, shown as dotted lines.
}
\label{fig.unification}
\end{figure}

There are many scenarios for realizing $\alpha_2(\Mp)\sim 2\pi$ (so that $f\simeq \Mp$).
For example, we can simply include $3$ $\SU(2)$ triplets at $\sim 10^7\, \textrm{GeV}$.
As another possibility consistent with a coupling unification as in grand unified theories (GUT), 
let us consider an $\SU(2)$ triplet and an $\SU(3)$ octet at $10^{12} \textrm{GeV}$.
We see that all gauge couplings meet at the Planck scale \cite{Bachas:1995yt,Bhattacharyya:2013xba}.
The value $\alpha_2(\Mp)\sim 2\pi$ can then be achieved by 
including $4$ pairs of GUT-like multiplets $\bf{5}$ and $\bar{\bf{5}}$ at $\sim 10\, \textrm{TeV}$ (see Figure \ref{fig.unification}).
In any of those scenarios, $\Lambda_a \sim \Lambda_0$ is not affected by the Supersymmetric Miracle.

Incidentally, in a supersymmetric theory the axion is complexified, so that we have 
another scalar (the saxion) \footnote{The existence of the saxion is predicted more generally from one of the swampland conjectures in \cite{Ooguri:2006in}.}.
The saxion causes a cosmological problem similar to the Polonyi 
problem \cite{Coughlan:1983ci,Goncharov:1984qm,Ellis:1986zt,Banks:1993en,deCarlos:1993wie}.
This problem can easily be solved if the saxion strongly couples to the inflaton \cite{Linde:1996cx,Nakayama:2012mf}.

Finally, let us discuss the issues of the global FN symmetry and the $R$-symmetry.
As the swampland conjecture precludes the existence of exact global symmetries,
those symmetries cannot be exact.
On the other hand, the sizes and the patterns of the explicit breaking due to the quantum gravity
are under debate.
Thus, for example, if the sizes of the explicit breaking are of ${\cal O}(10^{-1})$ or smaller for
unit charges, the successful features of the EW axion quintessence are not spoiled%
\footnote{Without any global symmetry, it is not possible to suppress a local operator which
can absorb all the fermion zero modes associated with the electroweak instanton. 
Such an operator can change the estimation of $\Lambda_a$ in Eq.\,(\ref{Lambda_a_SUSY}) drastically.
Once we allow the global $R$-symmetry, on the other hand, such a local operator is 
highly suppressed even if it is an approximate symmetry.
}.

In summary, in this Letter we pointed out that the swampland conjectures 
 bring a dynamical axion for the electroweak $\SU(2)$ gauge theory 
as an interesting candidate for the quintessence field. 
We also discussed constraints both from the weak gravity conjecture, from the conjecture of no global symmetry, 
and from observations.
We found that those constraints are satisfied elegantly in a supersymmetric extension of the standard model.
There, the large decay constant, $f_a \sim\Mp$, is achieved without spoiling the successful relation, 
$\Lambda_a \sim \Lambda_0$.
It is a fascinating question to explore if our scenario can really be realized in a 
specific setup inside theories of quantum gravity, such as string theory.

\bigskip\noindent
{\bf Acknowledgements}

We would like to thank Masahiro Kawasaki for collaboration in the initial stages of this project.
This research was supported in part by WPI Research Center Initiative, MEXT, Japan (MI, MY and TTY),
and in part by JSPS Grant-in-Aid for Scientific Research No.\ 15H05889, No.\ 16H03991, No.\ 18H05542 (MI), No.\ 17KK0087 (MY), 
No.\ 26104001, No.\ 26104009, No.\ 16H02176 (TTY), and No.\ 17H02878 (MI and TTY).
TTY is a Hamamatsu Professor at Kavli IPMU.

\bibliographystyle{apsrev4-1}
\bibliography{swampland_Q}

\begin{thebibliography}{62}%
\makeatletter
\providecommand \@ifxundefined [1]{%
 \@ifx{#1\undefined}
}%
\providecommand \@ifnum [1]{%
 \ifnum #1\expandafter \@firstoftwo
 \else \expandafter \@secondoftwo
 \fi
}%
\providecommand \@ifx [1]{%
 \ifx #1\expandafter \@firstoftwo
 \else \expandafter \@secondoftwo
 \fi
}%
\providecommand \natexlab [1]{#1}%
\providecommand \enquote  [1]{``#1''}%
\providecommand \bibnamefont  [1]{#1}%
\providecommand \bibfnamefont [1]{#1}%
\providecommand \citenamefont [1]{#1}%
\providecommand \href@noop [0]{\@secondoftwo}%
\providecommand \href [0]{\begingroup \@sanitize@url \@href}%
\providecommand \@href[1]{\@@startlink{#1}\@@href}%
\providecommand \@@href[1]{\endgroup#1\@@endlink}%
\providecommand \@sanitize@url [0]{\catcode `\\12\catcode `\$12\catcode
  `\&12\catcode `\#12\catcode `\^12\catcode `\_12\catcode `\%12\relax}%
\providecommand \@@startlink[1]{}%
\providecommand \@@endlink[0]{}%
\providecommand \url  [0]{\begingroup\@sanitize@url \@url }%
\providecommand \@url [1]{\endgroup\@href {#1}{\urlprefix }}%
\providecommand \urlprefix  [0]{URL }%
\providecommand \Eprint [0]{\href }%
\providecommand \doibase [0]{http://dx.doi.org/}%
\providecommand \selectlanguage [0]{\@gobble}%
\providecommand \bibinfo  [0]{\@secondoftwo}%
\providecommand \bibfield  [0]{\@secondoftwo}%
\providecommand \translation [1]{[#1]}%
\providecommand \BibitemOpen [0]{}%
\providecommand \bibitemStop [0]{}%
\providecommand \bibitemNoStop [0]{.\EOS\space}%
\providecommand \EOS [0]{\spacefactor3000\relax}%
\providecommand \BibitemShut  [1]{\csname bibitem#1\endcsname}%
\let\auto@bib@innerbib\@empty
\bibitem [{\citenamefont {Vafa}(2005)}]{Vafa:2005ui}%
  \BibitemOpen
  \bibfield  {author} {\bibinfo {author} {\bibfnamefont {C.}~\bibnamefont
  {Vafa}},\ }\href@noop {} {\  (\bibinfo {year} {2005})},\ \Eprint
  {http://arxiv.org/abs/hep-th/0509212} {arXiv:hep-th/0509212 [hep-th]}
  \BibitemShut {NoStop}%
\bibitem [{\citenamefont {Ooguri}\ and\ \citenamefont
  {Vafa}(2007)}]{Ooguri:2006in}%
  \BibitemOpen
  \bibfield  {author} {\bibinfo {author} {\bibfnamefont {H.}~\bibnamefont
  {Ooguri}}\ and\ \bibinfo {author} {\bibfnamefont {C.}~\bibnamefont {Vafa}},\
  }\href {\doibase 10.1016/j.nuclphysb.2006.10.033} {\bibfield  {journal}
  {\bibinfo  {journal} {Nucl. Phys.}\ }\textbf {\bibinfo {volume} {B766}},\
  \bibinfo {pages} {21} (\bibinfo {year} {2007})},\ \Eprint
  {http://arxiv.org/abs/hep-th/0605264} {arXiv:hep-th/0605264 [hep-th]}
  \BibitemShut {NoStop}%
\bibitem [{\citenamefont {Brennan}\ \emph {et~al.}(2017)\citenamefont
  {Brennan}, \citenamefont {Carta},\ and\ \citenamefont
  {Vafa}}]{Brennan:2017rbf}%
  \BibitemOpen
  \bibfield  {author} {\bibinfo {author} {\bibfnamefont {T.~D.}\ \bibnamefont
  {Brennan}}, \bibinfo {author} {\bibfnamefont {F.}~\bibnamefont {Carta}}, \
  and\ \bibinfo {author} {\bibfnamefont {C.}~\bibnamefont {Vafa}},\ }\bibfield
  {booktitle} {\emph {\bibinfo {booktitle} {{Proceedings, Theoretical Advanced
  Study Institute in Elementary Particle Physics: Physics at the Fundamental
  Frontier (TASI 2017): Boulder, CO, USA, June 5-30, 2017}}},\ }\href {\doibase
  10.22323/1.305.0015} {\bibfield  {journal} {\bibinfo  {journal} {PoS}\
  }\textbf {\bibinfo {volume} {TASI2017}},\ \bibinfo {pages} {015} (\bibinfo
  {year} {2017})},\ \Eprint {http://arxiv.org/abs/1711.00864} {arXiv:1711.00864
  [hep-th]} \BibitemShut {NoStop}%
\bibitem [{\citenamefont {Dvali}\ and\ \citenamefont
  {Gomez}(2016)}]{Dvali:2014gua}%
  \BibitemOpen
  \bibfield  {author} {\bibinfo {author} {\bibfnamefont {G.}~\bibnamefont
  {Dvali}}\ and\ \bibinfo {author} {\bibfnamefont {C.}~\bibnamefont {Gomez}},\
  }\href {\doibase 10.1002/andp.201500216} {\bibfield  {journal} {\bibinfo
  {journal} {Annalen Phys.}\ }\textbf {\bibinfo {volume} {528}},\ \bibinfo
  {pages} {68} (\bibinfo {year} {2016})},\ \Eprint
  {http://arxiv.org/abs/1412.8077} {arXiv:1412.8077 [hep-th]} \BibitemShut
  {NoStop}%
\bibitem [{\citenamefont {Dvali}\ \emph {et~al.}(2017)\citenamefont {Dvali},
  \citenamefont {Gomez},\ and\ \citenamefont {Zell}}]{Dvali:2017eba}%
  \BibitemOpen
  \bibfield  {author} {\bibinfo {author} {\bibfnamefont {G.}~\bibnamefont
  {Dvali}}, \bibinfo {author} {\bibfnamefont {C.}~\bibnamefont {Gomez}}, \ and\
  \bibinfo {author} {\bibfnamefont {S.}~\bibnamefont {Zell}},\ }\href {\doibase
  10.1088/1475-7516/2017/06/028} {\bibfield  {journal} {\bibinfo  {journal}
  {JCAP}\ }\textbf {\bibinfo {volume} {1706}},\ \bibinfo {pages} {028}
  (\bibinfo {year} {2017})},\ \Eprint {http://arxiv.org/abs/1701.08776}
  {arXiv:1701.08776 [hep-th]} \BibitemShut {NoStop}%
\bibitem [{\citenamefont {Sethi}(2018)}]{Sethi:2017phn}%
  \BibitemOpen
  \bibfield  {author} {\bibinfo {author} {\bibfnamefont {S.}~\bibnamefont
  {Sethi}},\ }\href {\doibase 10.1007/JHEP10(2018)022} {\bibfield  {journal}
  {\bibinfo  {journal} {JHEP}\ }\textbf {\bibinfo {volume} {10}},\ \bibinfo
  {pages} {022} (\bibinfo {year} {2018})},\ \Eprint
  {http://arxiv.org/abs/1709.03554} {arXiv:1709.03554 [hep-th]} \BibitemShut
  {NoStop}%
\bibitem [{\citenamefont {Danielsson}\ and\ \citenamefont
  {Van~Riet}(2018)}]{Danielsson:2018ztv}%
  \BibitemOpen
  \bibfield  {author} {\bibinfo {author} {\bibfnamefont {U.~H.}\ \bibnamefont
  {Danielsson}}\ and\ \bibinfo {author} {\bibfnamefont {T.}~\bibnamefont
  {Van~Riet}},\ }\href {\doibase 10.1142/S0218271818300070} {\bibfield
  {journal} {\bibinfo  {journal} {Int. J. Mod. Phys.}\ }\textbf {\bibinfo
  {volume} {D27}},\ \bibinfo {pages} {1830007} (\bibinfo {year} {2018})},\
  \Eprint {http://arxiv.org/abs/1804.01120} {arXiv:1804.01120 [hep-th]}
  \BibitemShut {NoStop}%
\bibitem [{\citenamefont {Obied}\ \emph {et~al.}(2018)\citenamefont {Obied},
  \citenamefont {Ooguri}, \citenamefont {Spodyneiko},\ and\ \citenamefont
  {Vafa}}]{Obied:2018sgi}%
  \BibitemOpen
  \bibfield  {author} {\bibinfo {author} {\bibfnamefont {G.}~\bibnamefont
  {Obied}}, \bibinfo {author} {\bibfnamefont {H.}~\bibnamefont {Ooguri}},
  \bibinfo {author} {\bibfnamefont {L.}~\bibnamefont {Spodyneiko}}, \ and\
  \bibinfo {author} {\bibfnamefont {C.}~\bibnamefont {Vafa}},\ }\href@noop {}
  {\  (\bibinfo {year} {2018})},\ \Eprint {http://arxiv.org/abs/1806.08362}
  {arXiv:1806.08362 [hep-th]} \BibitemShut {NoStop}%
\bibitem [{\citenamefont {Andriot}(2018)}]{Andriot:2018wzk}%
  \BibitemOpen
  \bibfield  {author} {\bibinfo {author} {\bibfnamefont {D.}~\bibnamefont
  {Andriot}},\ }\href {\doibase 10.1016/j.physletb.2018.09.022} {\bibfield
  {journal} {\bibinfo  {journal} {Phys. Lett.}\ }\textbf {\bibinfo {volume}
  {B785}},\ \bibinfo {pages} {570} (\bibinfo {year} {2018})},\ \Eprint
  {http://arxiv.org/abs/1806.10999} {arXiv:1806.10999 [hep-th]} \BibitemShut
  {NoStop}%
\bibitem [{\citenamefont {Garg}\ and\ \citenamefont
  {Krishnan}(2018)}]{Garg:2018reu}%
  \BibitemOpen
  \bibfield  {author} {\bibinfo {author} {\bibfnamefont {S.~K.}\ \bibnamefont
  {Garg}}\ and\ \bibinfo {author} {\bibfnamefont {C.}~\bibnamefont
  {Krishnan}},\ }\href@noop {} {\  (\bibinfo {year} {2018})},\ \Eprint
  {http://arxiv.org/abs/1807.05193} {arXiv:1807.05193 [hep-th]} \BibitemShut
  {NoStop}%
\bibitem [{\citenamefont {Murayama}\ \emph {et~al.}(2018)\citenamefont
  {Murayama}, \citenamefont {Yamazaki},\ and\ \citenamefont
  {Yanagida}}]{Murayama:2018lie}%
  \BibitemOpen
  \bibfield  {author} {\bibinfo {author} {\bibfnamefont {H.}~\bibnamefont
  {Murayama}}, \bibinfo {author} {\bibfnamefont {M.}~\bibnamefont {Yamazaki}},
  \ and\ \bibinfo {author} {\bibfnamefont {T.~T.}\ \bibnamefont {Yanagida}},\
  }\href {\doibase 10.1007/JHEP12(2018)032} {\bibfield  {journal} {\bibinfo
  {journal} {JHEP}\ }\textbf {\bibinfo {volume} {12}},\ \bibinfo {pages} {032}
  (\bibinfo {year} {2018})},\ \Eprint {http://arxiv.org/abs/1809.00478}
  {arXiv:1809.00478 [hep-th]} \BibitemShut {NoStop}%
\bibitem [{\citenamefont {Ooguri}\ \emph {et~al.}(2019)\citenamefont {Ooguri},
  \citenamefont {Palti}, \citenamefont {Shiu},\ and\ \citenamefont
  {Vafa}}]{Ooguri:2018wrx}%
  \BibitemOpen
  \bibfield  {author} {\bibinfo {author} {\bibfnamefont {H.}~\bibnamefont
  {Ooguri}}, \bibinfo {author} {\bibfnamefont {E.}~\bibnamefont {Palti}},
  \bibinfo {author} {\bibfnamefont {G.}~\bibnamefont {Shiu}}, \ and\ \bibinfo
  {author} {\bibfnamefont {C.}~\bibnamefont {Vafa}},\ }\href {\doibase
  10.1016/j.physletb.2018.11.018} {\bibfield  {journal} {\bibinfo  {journal}
  {Phys. Lett.}\ }\textbf {\bibinfo {volume} {B788}},\ \bibinfo {pages} {180}
  (\bibinfo {year} {2019})},\ \Eprint {http://arxiv.org/abs/1810.05506}
  {arXiv:1810.05506 [hep-th]} \BibitemShut {NoStop}%
\bibitem [{\citenamefont {Garg}\ \emph {et~al.}(2019)\citenamefont {Garg},
  \citenamefont {Krishnan},\ and\ \citenamefont {Zaid~Zaz}}]{Garg:2018zdg}%
  \BibitemOpen
  \bibfield  {author} {\bibinfo {author} {\bibfnamefont {S.~K.}\ \bibnamefont
  {Garg}}, \bibinfo {author} {\bibfnamefont {C.}~\bibnamefont {Krishnan}}, \
  and\ \bibinfo {author} {\bibfnamefont {M.}~\bibnamefont {Zaid~Zaz}},\ }\href
  {\doibase 10.1007/JHEP03(2019)029} {\bibfield  {journal} {\bibinfo  {journal}
  {JHEP}\ }\textbf {\bibinfo {volume} {03}},\ \bibinfo {pages} {029} (\bibinfo
  {year} {2019})},\ \Eprint {http://arxiv.org/abs/1810.09406} {arXiv:1810.09406
  [hep-th]} \BibitemShut {NoStop}%
\bibitem [{\citenamefont {Maldacena}\ and\ \citenamefont
  {Nunez}(2001)}]{Maldacena:2000mw}%
  \BibitemOpen
  \bibfield  {author} {\bibinfo {author} {\bibfnamefont {J.~M.}\ \bibnamefont
  {Maldacena}}\ and\ \bibinfo {author} {\bibfnamefont {C.}~\bibnamefont
  {Nunez}},\ }\bibfield  {booktitle} {\emph {\bibinfo {booktitle}
  {{Superstrings. Proceedings, International Conference, Strings 2000, Ann
  Arbor, USA, July 10-15, 2000}}},\ }\href {\doibase 10.1142/S0217751X01003935,
  10.1142/S0217751X01003937} {\bibfield  {journal} {\bibinfo  {journal} {Int.
  J. Mod. Phys.}\ }\textbf {\bibinfo {volume} {A16}},\ \bibinfo {pages} {822}
  (\bibinfo {year} {2001})},\ \bibinfo {note} {[,182(2000)]},\ \Eprint
  {http://arxiv.org/abs/hep-th/0007018} {arXiv:hep-th/0007018 [hep-th]}
  \BibitemShut {NoStop}%
\bibitem [{\citenamefont {Klaewer}\ and\ \citenamefont
  {Palti}(2017)}]{Klaewer:2016kiy}%
  \BibitemOpen
  \bibfield  {author} {\bibinfo {author} {\bibfnamefont {D.}~\bibnamefont
  {Klaewer}}\ and\ \bibinfo {author} {\bibfnamefont {E.}~\bibnamefont
  {Palti}},\ }\href {\doibase 10.1007/JHEP01(2017)088} {\bibfield  {journal}
  {\bibinfo  {journal} {JHEP}\ }\textbf {\bibinfo {volume} {01}},\ \bibinfo
  {pages} {088} (\bibinfo {year} {2017})},\ \Eprint
  {http://arxiv.org/abs/1610.00010} {arXiv:1610.00010 [hep-th]} \BibitemShut
  {NoStop}%
\bibitem [{\citenamefont {Dine}\ and\ \citenamefont
  {Seiberg}(1985)}]{Dine:1985he}%
  \BibitemOpen
  \bibfield  {author} {\bibinfo {author} {\bibfnamefont {M.}~\bibnamefont
  {Dine}}\ and\ \bibinfo {author} {\bibfnamefont {N.}~\bibnamefont {Seiberg}},\
  }\href {\doibase 10.1016/0370-2693(85)90927-X} {\bibfield  {journal}
  {\bibinfo  {journal} {Phys. Lett.}\ }\textbf {\bibinfo {volume} {162B}},\
  \bibinfo {pages} {299} (\bibinfo {year} {1985})}\BibitemShut {NoStop}%
\bibitem [{\citenamefont {Hebecker}\ and\ \citenamefont
  {Wrase}(2019)}]{Hebecker:2018vxz}%
  \BibitemOpen
  \bibfield  {author} {\bibinfo {author} {\bibfnamefont {A.}~\bibnamefont
  {Hebecker}}\ and\ \bibinfo {author} {\bibfnamefont {T.}~\bibnamefont
  {Wrase}},\ }\href {\doibase 10.1002/prop.201800097} {\bibfield  {journal}
  {\bibinfo  {journal} {Fortsch. Phys.}\ }\textbf {\bibinfo {volume} {67}},\
  \bibinfo {pages} {1800097} (\bibinfo {year} {2019})},\ \Eprint
  {http://arxiv.org/abs/1810.08182} {arXiv:1810.08182 [hep-th]} \BibitemShut
  {NoStop}%
\bibitem [{\citenamefont {Ooguri}\ and\ \citenamefont
  {Vafa}(2017)}]{Ooguri:2016pdq}%
  \BibitemOpen
  \bibfield  {author} {\bibinfo {author} {\bibfnamefont {H.}~\bibnamefont
  {Ooguri}}\ and\ \bibinfo {author} {\bibfnamefont {C.}~\bibnamefont {Vafa}},\
  }\href {\doibase 10.4310/ATMP.2017.v21.n7.a8} {\bibfield  {journal} {\bibinfo
   {journal} {Adv. Theor. Math. Phys.}\ }\textbf {\bibinfo {volume} {21}},\
  \bibinfo {pages} {1787} (\bibinfo {year} {2017})},\ \Eprint
  {http://arxiv.org/abs/1610.01533} {arXiv:1610.01533 [hep-th]} \BibitemShut
  {NoStop}%
\bibitem [{\citenamefont {Callan}\ \emph {et~al.}(1976)\citenamefont {Callan},
  \citenamefont {Dashen},\ and\ \citenamefont {Gross}}]{Callan:1976je}%
  \BibitemOpen
  \bibfield  {author} {\bibinfo {author} {\bibfnamefont {C.~G.}\ \bibnamefont
  {Callan}, \bibfnamefont {Jr.}}, \bibinfo {author} {\bibfnamefont {R.~F.}\
  \bibnamefont {Dashen}}, \ and\ \bibinfo {author} {\bibfnamefont {D.~J.}\
  \bibnamefont {Gross}},\ }\href {\doibase 10.1016/0370-2693(76)90277-X}
  {\bibfield  {journal} {\bibinfo  {journal} {Phys. Lett.}\ }\textbf {\bibinfo
  {volume} {63B}},\ \bibinfo {pages} {334} (\bibinfo {year} {1976})},\ \bibinfo
  {note} {[,357(1976)]}\BibitemShut {NoStop}%
\bibitem [{\citenamefont {Misner}\ and\ \citenamefont
  {Wheeler}(1957)}]{Misner:1957mt}%
  \BibitemOpen
  \bibfield  {author} {\bibinfo {author} {\bibfnamefont {C.~W.}\ \bibnamefont
  {Misner}}\ and\ \bibinfo {author} {\bibfnamefont {J.~A.}\ \bibnamefont
  {Wheeler}},\ }\href {\doibase 10.1016/0003-4916(57)90049-0} {\bibfield
  {journal} {\bibinfo  {journal} {Annals Phys.}\ }\textbf {\bibinfo {volume}
  {2}},\ \bibinfo {pages} {525} (\bibinfo {year} {1957})}\BibitemShut {NoStop}%
\bibitem [{\citenamefont {Polchinski}(2004)}]{Polchinski:2003bq}%
  \BibitemOpen
  \bibfield  {author} {\bibinfo {author} {\bibfnamefont {J.}~\bibnamefont
  {Polchinski}},\ }\bibfield  {booktitle} {\emph {\bibinfo {booktitle}
  {{Proceedings, Dirac Centennial Symposium, Tallahassee, USA, December 6-7,
  2002}}},\ }\href {\doibase 10.1142/S0217751X0401866X} {\bibfield  {journal}
  {\bibinfo  {journal} {Int. J. Mod. Phys.}\ }\textbf {\bibinfo {volume}
  {A19S1}},\ \bibinfo {pages} {145} (\bibinfo {year} {2004})},\ \bibinfo {note}
  {[,145(2003)]},\ \Eprint {http://arxiv.org/abs/hep-th/0304042}
  {arXiv:hep-th/0304042 [hep-th]} \BibitemShut {NoStop}%
\bibitem [{\citenamefont {Banks}\ and\ \citenamefont
  {Seiberg}(2011)}]{Banks:2010zn}%
  \BibitemOpen
  \bibfield  {author} {\bibinfo {author} {\bibfnamefont {T.}~\bibnamefont
  {Banks}}\ and\ \bibinfo {author} {\bibfnamefont {N.}~\bibnamefont
  {Seiberg}},\ }\href {\doibase 10.1103/PhysRevD.83.084019} {\bibfield
  {journal} {\bibinfo  {journal} {Phys. Rev.}\ }\textbf {\bibinfo {volume}
  {D83}},\ \bibinfo {pages} {084019} (\bibinfo {year} {2011})},\ \Eprint
  {http://arxiv.org/abs/1011.5120} {arXiv:1011.5120 [hep-th]} \BibitemShut
  {NoStop}%
\bibitem [{\citenamefont {Harlow}\ and\ \citenamefont
  {Ooguri}(2018)}]{Harlow:2018tng}%
  \BibitemOpen
  \bibfield  {author} {\bibinfo {author} {\bibfnamefont {D.}~\bibnamefont
  {Harlow}}\ and\ \bibinfo {author} {\bibfnamefont {H.}~\bibnamefont
  {Ooguri}},\ }\href@noop {} {\  (\bibinfo {year} {2018})},\ \Eprint
  {http://arxiv.org/abs/1810.05338} {arXiv:1810.05338 [hep-th]} \BibitemShut
  {NoStop}%
\bibitem [{\citenamefont {Ibanez}\ \emph {et~al.}(2017)\citenamefont {Ibanez},
  \citenamefont {Martin-Lozano},\ and\ \citenamefont
  {Valenzuela}}]{Ibanez:2017kvh}%
  \BibitemOpen
  \bibfield  {author} {\bibinfo {author} {\bibfnamefont {L.~E.}\ \bibnamefont
  {Ibanez}}, \bibinfo {author} {\bibfnamefont {V.}~\bibnamefont
  {Martin-Lozano}}, \ and\ \bibinfo {author} {\bibfnamefont {I.}~\bibnamefont
  {Valenzuela}},\ }\href {\doibase 10.1007/JHEP11(2017)066} {\bibfield
  {journal} {\bibinfo  {journal} {JHEP}\ }\textbf {\bibinfo {volume} {11}},\
  \bibinfo {pages} {066} (\bibinfo {year} {2017})},\ \Eprint
  {http://arxiv.org/abs/1706.05392} {arXiv:1706.05392 [hep-th]} \BibitemShut
  {NoStop}%
\bibitem [{\citenamefont {'t~Hooft}(1976)}]{tHooft:1976snw}%
  \BibitemOpen
  \bibfield  {author} {\bibinfo {author} {\bibfnamefont {G.}~\bibnamefont
  {'t~Hooft}},\ }\href {\doibase 10.1103/PhysRevD.18.2199.3,
  10.1103/PhysRevD.14.3432} {\bibfield  {journal} {\bibinfo  {journal} {Phys.
  Rev.}\ }\textbf {\bibinfo {volume} {D14}},\ \bibinfo {pages} {3432} (\bibinfo
  {year} {1976})},\ \bibinfo {note} {[,70(1976)]}\BibitemShut {NoStop}%
\bibitem [{\citenamefont {Denef}\ \emph {et~al.}(2018)\citenamefont {Denef},
  \citenamefont {Hebecker},\ and\ \citenamefont {Wrase}}]{Denef:2018etk}%
  \BibitemOpen
  \bibfield  {author} {\bibinfo {author} {\bibfnamefont {F.}~\bibnamefont
  {Denef}}, \bibinfo {author} {\bibfnamefont {A.}~\bibnamefont {Hebecker}}, \
  and\ \bibinfo {author} {\bibfnamefont {T.}~\bibnamefont {Wrase}},\ }\href
  {\doibase 10.1103/PhysRevD.98.086004} {\bibfield  {journal} {\bibinfo
  {journal} {Phys. Rev.}\ }\textbf {\bibinfo {volume} {D98}},\ \bibinfo {pages}
  {086004} (\bibinfo {year} {2018})},\ \Eprint
  {http://arxiv.org/abs/1807.06581} {arXiv:1807.06581 [hep-th]} \BibitemShut
  {NoStop}%
\bibitem [{\citenamefont {Conlon}(2018)}]{Conlon:2018eyr}%
  \BibitemOpen
  \bibfield  {author} {\bibinfo {author} {\bibfnamefont {J.~P.}\ \bibnamefont
  {Conlon}},\ }\href {\doibase 10.1142/S0217751X18501786} {\bibfield  {journal}
  {\bibinfo  {journal} {Int. J. Mod. Phys.}\ }\textbf {\bibinfo {volume}
  {A33}},\ \bibinfo {pages} {1850178} (\bibinfo {year} {2018})},\ \Eprint
  {http://arxiv.org/abs/1808.05040} {arXiv:1808.05040 [hep-th]} \BibitemShut
  {NoStop}%
\bibitem [{\citenamefont {Choi}\ \emph {et~al.}(2018)\citenamefont {Choi},
  \citenamefont {Chway},\ and\ \citenamefont {Shin}}]{Choi:2018rze}%
  \BibitemOpen
  \bibfield  {author} {\bibinfo {author} {\bibfnamefont {K.}~\bibnamefont
  {Choi}}, \bibinfo {author} {\bibfnamefont {D.}~\bibnamefont {Chway}}, \ and\
  \bibinfo {author} {\bibfnamefont {C.~S.}\ \bibnamefont {Shin}},\ }\href
  {\doibase 10.1007/JHEP11(2018)142} {\bibfield  {journal} {\bibinfo  {journal}
  {JHEP}\ }\textbf {\bibinfo {volume} {11}},\ \bibinfo {pages} {142} (\bibinfo
  {year} {2018})},\ \Eprint {http://arxiv.org/abs/1809.01475} {arXiv:1809.01475
  [hep-th]} \BibitemShut {NoStop}%
\bibitem [{\citenamefont {Hamaguchi}\ \emph {et~al.}(2018)\citenamefont
  {Hamaguchi}, \citenamefont {Ibe},\ and\ \citenamefont
  {Moroi}}]{Hamaguchi:2018vtv}%
  \BibitemOpen
  \bibfield  {author} {\bibinfo {author} {\bibfnamefont {K.}~\bibnamefont
  {Hamaguchi}}, \bibinfo {author} {\bibfnamefont {M.}~\bibnamefont {Ibe}}, \
  and\ \bibinfo {author} {\bibfnamefont {T.}~\bibnamefont {Moroi}},\ }\href
  {\doibase 10.1007/JHEP12(2018)023} {\bibfield  {journal} {\bibinfo  {journal}
  {JHEP}\ }\textbf {\bibinfo {volume} {12}},\ \bibinfo {pages} {023} (\bibinfo
  {year} {2018})},\ \Eprint {http://arxiv.org/abs/1810.02095} {arXiv:1810.02095
  [hep-th]} \BibitemShut {NoStop}%
\bibitem [{\citenamefont {Nomura}\ \emph {et~al.}(2000)\citenamefont {Nomura},
  \citenamefont {Watari},\ and\ \citenamefont {Yanagida}}]{Nomura:2000yk}%
  \BibitemOpen
  \bibfield  {author} {\bibinfo {author} {\bibfnamefont {Y.}~\bibnamefont
  {Nomura}}, \bibinfo {author} {\bibfnamefont {T.}~\bibnamefont {Watari}}, \
  and\ \bibinfo {author} {\bibfnamefont {T.}~\bibnamefont {Yanagida}},\ }\href
  {\doibase 10.1016/S0370-2693(00)00605-5} {\bibfield  {journal} {\bibinfo
  {journal} {Phys. Lett.}\ }\textbf {\bibinfo {volume} {B484}},\ \bibinfo
  {pages} {103} (\bibinfo {year} {2000})},\ \Eprint
  {http://arxiv.org/abs/hep-ph/0004182} {arXiv:hep-ph/0004182 [hep-ph]}
  \BibitemShut {NoStop}%
\bibitem [{\citenamefont {McLerran}\ \emph {et~al.}(2012)\citenamefont
  {McLerran}, \citenamefont {Pisarski},\ and\ \citenamefont
  {Skokov}}]{McLerran:2012mm}%
  \BibitemOpen
  \bibfield  {author} {\bibinfo {author} {\bibfnamefont {L.}~\bibnamefont
  {McLerran}}, \bibinfo {author} {\bibfnamefont {R.}~\bibnamefont {Pisarski}},
  \ and\ \bibinfo {author} {\bibfnamefont {V.}~\bibnamefont {Skokov}},\ }\href
  {\doibase 10.1016/j.physletb.2012.05.057} {\bibfield  {journal} {\bibinfo
  {journal} {Phys. Lett.}\ }\textbf {\bibinfo {volume} {B713}},\ \bibinfo
  {pages} {301} (\bibinfo {year} {2012})},\ \Eprint
  {http://arxiv.org/abs/1204.2533} {arXiv:1204.2533 [hep-ph]} \BibitemShut
  {NoStop}%
\bibitem [{\citenamefont {Fukugita}\ and\ \citenamefont
  {Yanagida}(1994)}]{Fukugita:1994hq}%
  \BibitemOpen
  \bibfield  {author} {\bibinfo {author} {\bibfnamefont {M.}~\bibnamefont
  {Fukugita}}\ and\ \bibinfo {author} {\bibfnamefont {T.}~\bibnamefont
  {Yanagida}},\ }\href@noop {} {\  (\bibinfo {year} {1994})},\ \bibinfo {note}
  {preprint YITP-K-1098}\BibitemShut {NoStop}%
\bibitem [{\citenamefont {Frieman}\ \emph {et~al.}(1995)\citenamefont
  {Frieman}, \citenamefont {Hill}, \citenamefont {Stebbins},\ and\
  \citenamefont {Waga}}]{Frieman:1995pm}%
  \BibitemOpen
  \bibfield  {author} {\bibinfo {author} {\bibfnamefont {J.~A.}\ \bibnamefont
  {Frieman}}, \bibinfo {author} {\bibfnamefont {C.~T.}\ \bibnamefont {Hill}},
  \bibinfo {author} {\bibfnamefont {A.}~\bibnamefont {Stebbins}}, \ and\
  \bibinfo {author} {\bibfnamefont {I.}~\bibnamefont {Waga}},\ }\href {\doibase
  10.1103/PhysRevLett.75.2077} {\bibfield  {journal} {\bibinfo  {journal}
  {Phys. Rev. Lett.}\ }\textbf {\bibinfo {volume} {75}},\ \bibinfo {pages}
  {2077} (\bibinfo {year} {1995})},\ \Eprint
  {http://arxiv.org/abs/astro-ph/9505060} {arXiv:astro-ph/9505060 [astro-ph]}
  \BibitemShut {NoStop}%
\bibitem [{\citenamefont {Choi}(2000)}]{Choi:1999xn}%
  \BibitemOpen
  \bibfield  {author} {\bibinfo {author} {\bibfnamefont {K.}~\bibnamefont
  {Choi}},\ }\href {\doibase 10.1103/PhysRevD.62.043509} {\bibfield  {journal}
  {\bibinfo  {journal} {Phys. Rev.}\ }\textbf {\bibinfo {volume} {D62}},\
  \bibinfo {pages} {043509} (\bibinfo {year} {2000})},\ \Eprint
  {http://arxiv.org/abs/hep-ph/9902292} {arXiv:hep-ph/9902292 [hep-ph]}
  \BibitemShut {NoStop}%
\bibitem [{\citenamefont {Ratra}\ and\ \citenamefont
  {Peebles}(1988)}]{Ratra:1987rm}%
  \BibitemOpen
  \bibfield  {author} {\bibinfo {author} {\bibfnamefont {B.}~\bibnamefont
  {Ratra}}\ and\ \bibinfo {author} {\bibfnamefont {P.~J.~E.}\ \bibnamefont
  {Peebles}},\ }\href {\doibase 10.1103/PhysRevD.37.3406} {\bibfield  {journal}
  {\bibinfo  {journal} {Phys. Rev.}\ }\textbf {\bibinfo {volume} {D37}},\
  \bibinfo {pages} {3406} (\bibinfo {year} {1988})}\BibitemShut {NoStop}%
\bibitem [{\citenamefont {Wetterich}(1988)}]{Wetterich:1987fm}%
  \BibitemOpen
  \bibfield  {author} {\bibinfo {author} {\bibfnamefont {C.}~\bibnamefont
  {Wetterich}},\ }\href {\doibase 10.1016/0550-3213(88)90193-9} {\bibfield
  {journal} {\bibinfo  {journal} {Nucl. Phys.}\ }\textbf {\bibinfo {volume}
  {B302}},\ \bibinfo {pages} {668} (\bibinfo {year} {1988})},\ \Eprint
  {http://arxiv.org/abs/1711.03844} {arXiv:1711.03844 [hep-th]} \BibitemShut
  {NoStop}%
\bibitem [{\citenamefont {Zlatev}\ \emph {et~al.}(1999)\citenamefont {Zlatev},
  \citenamefont {Wang},\ and\ \citenamefont {Steinhardt}}]{Zlatev:1998tr}%
  \BibitemOpen
  \bibfield  {author} {\bibinfo {author} {\bibfnamefont {I.}~\bibnamefont
  {Zlatev}}, \bibinfo {author} {\bibfnamefont {L.-M.}\ \bibnamefont {Wang}}, \
  and\ \bibinfo {author} {\bibfnamefont {P.~J.}\ \bibnamefont {Steinhardt}},\
  }\href {\doibase 10.1103/PhysRevLett.82.896} {\bibfield  {journal} {\bibinfo
  {journal} {Phys. Rev. Lett.}\ }\textbf {\bibinfo {volume} {82}},\ \bibinfo
  {pages} {896} (\bibinfo {year} {1999})},\ \Eprint
  {http://arxiv.org/abs/astro-ph/9807002} {arXiv:astro-ph/9807002 [astro-ph]}
  \BibitemShut {NoStop}%
\bibitem [{\citenamefont {Heisenberg}\ \emph {et~al.}(2018)\citenamefont
  {Heisenberg}, \citenamefont {Bartelmann}, \citenamefont {Brandenberger},\
  and\ \citenamefont {Refregier}}]{Heisenberg:2018yae}%
  \BibitemOpen
  \bibfield  {author} {\bibinfo {author} {\bibfnamefont {L.}~\bibnamefont
  {Heisenberg}}, \bibinfo {author} {\bibfnamefont {M.}~\bibnamefont
  {Bartelmann}}, \bibinfo {author} {\bibfnamefont {R.}~\bibnamefont
  {Brandenberger}}, \ and\ \bibinfo {author} {\bibfnamefont {A.}~\bibnamefont
  {Refregier}},\ }\href {\doibase 10.1103/PhysRevD.98.123502} {\bibfield
  {journal} {\bibinfo  {journal} {Phys. Rev.}\ }\textbf {\bibinfo {volume}
  {D98}},\ \bibinfo {pages} {123502} (\bibinfo {year} {2018})},\ \Eprint
  {http://arxiv.org/abs/1808.02877} {arXiv:1808.02877 [astro-ph.CO]}
  \BibitemShut {NoStop}%
\bibitem [{\citenamefont {Heisenberg}\ \emph {et~al.}(2019)\citenamefont
  {Heisenberg}, \citenamefont {Bartelmann}, \citenamefont {Brandenberger},\
  and\ \citenamefont {Refregier}}]{Heisenberg:2018rdu}%
  \BibitemOpen
  \bibfield  {author} {\bibinfo {author} {\bibfnamefont {L.}~\bibnamefont
  {Heisenberg}}, \bibinfo {author} {\bibfnamefont {M.}~\bibnamefont
  {Bartelmann}}, \bibinfo {author} {\bibfnamefont {R.}~\bibnamefont
  {Brandenberger}}, \ and\ \bibinfo {author} {\bibfnamefont {A.}~\bibnamefont
  {Refregier}},\ }\href {\doibase 10.1007/s11433-019-9392-7} {\bibfield
  {journal} {\bibinfo  {journal} {Sci. China Phys. Mech. Astron.}\ }\textbf
  {\bibinfo {volume} {62}},\ \bibinfo {pages} {990421} (\bibinfo {year}
  {2019})},\ \Eprint {http://arxiv.org/abs/1809.00154} {arXiv:1809.00154
  [astro-ph.CO]} \BibitemShut {NoStop}%
\bibitem [{\citenamefont {D'Amico}\ \emph {et~al.}(2018)\citenamefont
  {D'Amico}, \citenamefont {Kaloper},\ and\ \citenamefont
  {Lawrence}}]{DAmico:2018mnx}%
  \BibitemOpen
  \bibfield  {author} {\bibinfo {author} {\bibfnamefont {G.}~\bibnamefont
  {D'Amico}}, \bibinfo {author} {\bibfnamefont {N.}~\bibnamefont {Kaloper}}, \
  and\ \bibinfo {author} {\bibfnamefont {A.}~\bibnamefont {Lawrence}},\
  }\href@noop {} {\  (\bibinfo {year} {2018})},\ \Eprint
  {http://arxiv.org/abs/1809.05109} {arXiv:1809.05109 [hep-th]} \BibitemShut
  {NoStop}%
\bibitem [{\citenamefont {Agrawal}\ and\ \citenamefont
  {Obied}(2019)}]{Agrawal:2018rcg}%
  \BibitemOpen
  \bibfield  {author} {\bibinfo {author} {\bibfnamefont {P.}~\bibnamefont
  {Agrawal}}\ and\ \bibinfo {author} {\bibfnamefont {G.}~\bibnamefont
  {Obied}},\ }\href {\doibase 10.1007/JHEP06(2019)103} {\bibfield  {journal}
  {\bibinfo  {journal} {JHEP}\ }\textbf {\bibinfo {volume} {06}},\ \bibinfo
  {pages} {103} (\bibinfo {year} {2019})},\ \Eprint
  {http://arxiv.org/abs/1811.00554} {arXiv:1811.00554 [hep-ph]} \BibitemShut
  {NoStop}%
\bibitem [{\citenamefont {Arkani-Hamed}\ \emph {et~al.}(2007)\citenamefont
  {Arkani-Hamed}, \citenamefont {Motl}, \citenamefont {Nicolis},\ and\
  \citenamefont {Vafa}}]{ArkaniHamed:2006dz}%
  \BibitemOpen
  \bibfield  {author} {\bibinfo {author} {\bibfnamefont {N.}~\bibnamefont
  {Arkani-Hamed}}, \bibinfo {author} {\bibfnamefont {L.}~\bibnamefont {Motl}},
  \bibinfo {author} {\bibfnamefont {A.}~\bibnamefont {Nicolis}}, \ and\
  \bibinfo {author} {\bibfnamefont {C.}~\bibnamefont {Vafa}},\ }\href {\doibase
  10.1088/1126-6708/2007/06/060} {\bibfield  {journal} {\bibinfo  {journal}
  {JHEP}\ }\textbf {\bibinfo {volume} {06}},\ \bibinfo {pages} {060} (\bibinfo
  {year} {2007})},\ \Eprint {http://arxiv.org/abs/hep-th/0601001}
  {arXiv:hep-th/0601001 [hep-th]} \BibitemShut {NoStop}%
\bibitem [{\citenamefont {Dutta}\ and\ \citenamefont
  {Scherrer}(2008)}]{Dutta:2008qn}%
  \BibitemOpen
  \bibfield  {author} {\bibinfo {author} {\bibfnamefont {S.}~\bibnamefont
  {Dutta}}\ and\ \bibinfo {author} {\bibfnamefont {R.~J.}\ \bibnamefont
  {Scherrer}},\ }\href {\doibase 10.1103/PhysRevD.78.123525} {\bibfield
  {journal} {\bibinfo  {journal} {Phys. Rev.}\ }\textbf {\bibinfo {volume}
  {D78}},\ \bibinfo {pages} {123525} (\bibinfo {year} {2008})},\ \Eprint
  {http://arxiv.org/abs/0809.4441} {arXiv:0809.4441 [astro-ph]} \BibitemShut
  {NoStop}%
\bibitem [{\citenamefont {Sakai}\ and\ \citenamefont
  {Yanagida}(1982)}]{Sakai:1981pk}%
  \BibitemOpen
  \bibfield  {author} {\bibinfo {author} {\bibfnamefont {N.}~\bibnamefont
  {Sakai}}\ and\ \bibinfo {author} {\bibfnamefont {T.}~\bibnamefont
  {Yanagida}},\ }\href {\doibase 10.1016/0550-3213(82)90457-6} {\bibfield
  {journal} {\bibinfo  {journal} {Nucl. Phys.}\ }\textbf {\bibinfo {volume}
  {B197}},\ \bibinfo {pages} {533} (\bibinfo {year} {1982})}\BibitemShut
  {NoStop}%
\bibitem [{\citenamefont {Weinberg}(1982)}]{Weinberg:1981wj}%
  \BibitemOpen
  \bibfield  {author} {\bibinfo {author} {\bibfnamefont {S.}~\bibnamefont
  {Weinberg}},\ }\href {\doibase 10.1103/PhysRevD.26.287} {\bibfield  {journal}
  {\bibinfo  {journal} {Phys. Rev.}\ }\textbf {\bibinfo {volume} {D26}},\
  \bibinfo {pages} {287} (\bibinfo {year} {1982})}\BibitemShut {NoStop}%
\bibitem [{\citenamefont {Froggatt}\ and\ \citenamefont
  {Nielsen}(1979)}]{Froggatt:1978nt}%
  \BibitemOpen
  \bibfield  {author} {\bibinfo {author} {\bibfnamefont {C.~D.}\ \bibnamefont
  {Froggatt}}\ and\ \bibinfo {author} {\bibfnamefont {H.~B.}\ \bibnamefont
  {Nielsen}},\ }\href {\doibase 10.1016/0550-3213(79)90316-X} {\bibfield
  {journal} {\bibinfo  {journal} {Nucl. Phys.}\ }\textbf {\bibinfo {volume}
  {B147}},\ \bibinfo {pages} {277} (\bibinfo {year} {1979})}\BibitemShut
  {NoStop}%
\bibitem [{\citenamefont {Buchmuller}\ and\ \citenamefont
  {Yanagida}(1999)}]{Buchmuller:1998zf}%
  \BibitemOpen
  \bibfield  {author} {\bibinfo {author} {\bibfnamefont {W.}~\bibnamefont
  {Buchmuller}}\ and\ \bibinfo {author} {\bibfnamefont {T.}~\bibnamefont
  {Yanagida}},\ }\href {\doibase 10.1016/S0370-2693(98)01480-4} {\bibfield
  {journal} {\bibinfo  {journal} {Phys. Lett.}\ }\textbf {\bibinfo {volume}
  {B445}},\ \bibinfo {pages} {399} (\bibinfo {year} {1999})},\ \Eprint
  {http://arxiv.org/abs/hep-ph/9810308} {arXiv:hep-ph/9810308 [hep-ph]}
  \BibitemShut {NoStop}%
\bibitem [{\citenamefont {Bachas}\ \emph {et~al.}(1996)\citenamefont {Bachas},
  \citenamefont {Fabre},\ and\ \citenamefont {Yanagida}}]{Bachas:1995yt}%
  \BibitemOpen
  \bibfield  {author} {\bibinfo {author} {\bibfnamefont {C.}~\bibnamefont
  {Bachas}}, \bibinfo {author} {\bibfnamefont {C.}~\bibnamefont {Fabre}}, \
  and\ \bibinfo {author} {\bibfnamefont {T.}~\bibnamefont {Yanagida}},\ }\href
  {\doibase 10.1016/0370-2693(95)01561-2} {\bibfield  {journal} {\bibinfo
  {journal} {Phys. Lett.}\ }\textbf {\bibinfo {volume} {B370}},\ \bibinfo
  {pages} {49} (\bibinfo {year} {1996})},\ \Eprint
  {http://arxiv.org/abs/hep-th/9510094} {arXiv:hep-th/9510094 [hep-th]}
  \BibitemShut {NoStop}%
\bibitem [{\citenamefont {Bhattacharyya}\ \emph {et~al.}(2013)\citenamefont
  {Bhattacharyya}, \citenamefont {Bhattacherjee}, \citenamefont {Yanagida},\
  and\ \citenamefont {Yokozaki}}]{Bhattacharyya:2013xba}%
  \BibitemOpen
  \bibfield  {author} {\bibinfo {author} {\bibfnamefont {G.}~\bibnamefont
  {Bhattacharyya}}, \bibinfo {author} {\bibfnamefont {B.}~\bibnamefont
  {Bhattacherjee}}, \bibinfo {author} {\bibfnamefont {T.~T.}\ \bibnamefont
  {Yanagida}}, \ and\ \bibinfo {author} {\bibfnamefont {N.}~\bibnamefont
  {Yokozaki}},\ }\href {\doibase 10.1016/j.physletb.2013.07.040} {\bibfield
  {journal} {\bibinfo  {journal} {Phys. Lett.}\ }\textbf {\bibinfo {volume}
  {B725}},\ \bibinfo {pages} {339} (\bibinfo {year} {2013})},\ \Eprint
  {http://arxiv.org/abs/1304.2508} {arXiv:1304.2508 [hep-ph]} \BibitemShut
  {NoStop}%
\bibitem [{\citenamefont {Coughlan}\ \emph {et~al.}(1983)\citenamefont
  {Coughlan}, \citenamefont {Fischler}, \citenamefont {Kolb}, \citenamefont
  {Raby},\ and\ \citenamefont {Ross}}]{Coughlan:1983ci}%
  \BibitemOpen
  \bibfield  {author} {\bibinfo {author} {\bibfnamefont {G.~D.}\ \bibnamefont
  {Coughlan}}, \bibinfo {author} {\bibfnamefont {W.}~\bibnamefont {Fischler}},
  \bibinfo {author} {\bibfnamefont {E.~W.}\ \bibnamefont {Kolb}}, \bibinfo
  {author} {\bibfnamefont {S.}~\bibnamefont {Raby}}, \ and\ \bibinfo {author}
  {\bibfnamefont {G.~G.}\ \bibnamefont {Ross}},\ }\href {\doibase
  10.1016/0370-2693(83)91091-2} {\bibfield  {journal} {\bibinfo  {journal}
  {Phys. Lett.}\ }\textbf {\bibinfo {volume} {131B}},\ \bibinfo {pages} {59}
  (\bibinfo {year} {1983})}\BibitemShut {NoStop}%
\bibitem [{\citenamefont {Goncharov}\ \emph {et~al.}(1984)\citenamefont
  {Goncharov}, \citenamefont {Linde},\ and\ \citenamefont
  {Vysotsky}}]{Goncharov:1984qm}%
  \BibitemOpen
  \bibfield  {author} {\bibinfo {author} {\bibfnamefont {A.~S.}\ \bibnamefont
  {Goncharov}}, \bibinfo {author} {\bibfnamefont {A.~D.}\ \bibnamefont
  {Linde}}, \ and\ \bibinfo {author} {\bibfnamefont {M.~I.}\ \bibnamefont
  {Vysotsky}},\ }\href {\doibase 10.1016/0370-2693(84)90116-3} {\bibfield
  {journal} {\bibinfo  {journal} {Phys. Lett.}\ }\textbf {\bibinfo {volume}
  {147B}},\ \bibinfo {pages} {279} (\bibinfo {year} {1984})}\BibitemShut
  {NoStop}%
\bibitem [{\citenamefont {Ellis}\ \emph {et~al.}(1986)\citenamefont {Ellis},
  \citenamefont {Nanopoulos},\ and\ \citenamefont {Quiros}}]{Ellis:1986zt}%
  \BibitemOpen
  \bibfield  {author} {\bibinfo {author} {\bibfnamefont {J.~R.}\ \bibnamefont
  {Ellis}}, \bibinfo {author} {\bibfnamefont {D.~V.}\ \bibnamefont
  {Nanopoulos}}, \ and\ \bibinfo {author} {\bibfnamefont {M.}~\bibnamefont
  {Quiros}},\ }\href {\doibase 10.1016/0370-2693(86)90736-7} {\bibfield
  {journal} {\bibinfo  {journal} {Phys. Lett.}\ }\textbf {\bibinfo {volume}
  {B174}},\ \bibinfo {pages} {176} (\bibinfo {year} {1986})}\BibitemShut
  {NoStop}%
\bibitem [{\citenamefont {Banks}\ \emph {et~al.}(1994)\citenamefont {Banks},
  \citenamefont {Kaplan},\ and\ \citenamefont {Nelson}}]{Banks:1993en}%
  \BibitemOpen
  \bibfield  {author} {\bibinfo {author} {\bibfnamefont {T.}~\bibnamefont
  {Banks}}, \bibinfo {author} {\bibfnamefont {D.~B.}\ \bibnamefont {Kaplan}}, \
  and\ \bibinfo {author} {\bibfnamefont {A.~E.}\ \bibnamefont {Nelson}},\
  }\href {\doibase 10.1103/PhysRevD.49.779} {\bibfield  {journal} {\bibinfo
  {journal} {Phys. Rev.}\ }\textbf {\bibinfo {volume} {D49}},\ \bibinfo {pages}
  {779} (\bibinfo {year} {1994})},\ \Eprint
  {http://arxiv.org/abs/hep-ph/9308292} {arXiv:hep-ph/9308292 [hep-ph]}
  \BibitemShut {NoStop}%
\bibitem [{\citenamefont {de~Carlos}\ \emph {et~al.}(1993)\citenamefont
  {de~Carlos}, \citenamefont {Casas}, \citenamefont {Quevedo},\ and\
  \citenamefont {Roulet}}]{deCarlos:1993wie}%
  \BibitemOpen
  \bibfield  {author} {\bibinfo {author} {\bibfnamefont {B.}~\bibnamefont
  {de~Carlos}}, \bibinfo {author} {\bibfnamefont {J.~A.}\ \bibnamefont
  {Casas}}, \bibinfo {author} {\bibfnamefont {F.}~\bibnamefont {Quevedo}}, \
  and\ \bibinfo {author} {\bibfnamefont {E.}~\bibnamefont {Roulet}},\ }\href
  {\doibase 10.1016/0370-2693(93)91538-X} {\bibfield  {journal} {\bibinfo
  {journal} {Phys. Lett.}\ }\textbf {\bibinfo {volume} {B318}},\ \bibinfo
  {pages} {447} (\bibinfo {year} {1993})},\ \Eprint
  {http://arxiv.org/abs/hep-ph/9308325} {arXiv:hep-ph/9308325 [hep-ph]}
  \BibitemShut {NoStop}%
\bibitem [{\citenamefont {Linde}(1996)}]{Linde:1996cx}%
  \BibitemOpen
  \bibfield  {author} {\bibinfo {author} {\bibfnamefont {A.~D.}\ \bibnamefont
  {Linde}},\ }\href {\doibase 10.1103/PhysRevD.53.R4129} {\bibfield  {journal}
  {\bibinfo  {journal} {Phys. Rev.}\ }\textbf {\bibinfo {volume} {D53}},\
  \bibinfo {pages} {R4129} (\bibinfo {year} {1996})},\ \Eprint
  {http://arxiv.org/abs/hep-th/9601083} {arXiv:hep-th/9601083 [hep-th]}
  \BibitemShut {NoStop}%
\bibitem [{\citenamefont {Nakayama}\ \emph {et~al.}(2012)\citenamefont
  {Nakayama}, \citenamefont {Takahashi},\ and\ \citenamefont
  {Yanagida}}]{Nakayama:2012mf}%
  \BibitemOpen
  \bibfield  {author} {\bibinfo {author} {\bibfnamefont {K.}~\bibnamefont
  {Nakayama}}, \bibinfo {author} {\bibfnamefont {F.}~\bibnamefont {Takahashi}},
  \ and\ \bibinfo {author} {\bibfnamefont {T.~T.}\ \bibnamefont {Yanagida}},\
  }\href {\doibase 10.1016/j.physletb.2012.06.072} {\bibfield  {journal}
  {\bibinfo  {journal} {Phys. Lett.}\ }\textbf {\bibinfo {volume} {B714}},\
  \bibinfo {pages} {256} (\bibinfo {year} {2012})},\ \Eprint
  {http://arxiv.org/abs/1203.2085} {arXiv:1203.2085 [hep-ph]} \BibitemShut
  {NoStop}%
\bibitem [{\citenamefont {Peccei}\ and\ \citenamefont
  {Quinn}(1977)}]{Peccei:1977ur}%
  \BibitemOpen
  \bibfield  {author} {\bibinfo {author} {\bibfnamefont {R.~D.}\ \bibnamefont
  {Peccei}}\ and\ \bibinfo {author} {\bibfnamefont {H.~R.}\ \bibnamefont
  {Quinn}},\ }\href {\doibase 10.1103/PhysRevD.16.1791} {\bibfield  {journal}
  {\bibinfo  {journal} {Phys. Rev.}\ }\textbf {\bibinfo {volume} {D16}},\
  \bibinfo {pages} {1791} (\bibinfo {year} {1977})}\BibitemShut {NoStop}%
\bibitem [{\citenamefont {Weinberg}(1978)}]{Weinberg:1977ma}%
  \BibitemOpen
  \bibfield  {author} {\bibinfo {author} {\bibfnamefont {S.}~\bibnamefont
  {Weinberg}},\ }\href {\doibase 10.1103/PhysRevLett.40.223} {\bibfield
  {journal} {\bibinfo  {journal} {Phys. Rev. Lett.}\ }\textbf {\bibinfo
  {volume} {40}},\ \bibinfo {pages} {223} (\bibinfo {year} {1978})}\BibitemShut
  {NoStop}%
\bibitem [{\citenamefont {Wilczek}(1978)}]{Wilczek:1977pj}%
  \BibitemOpen
  \bibfield  {author} {\bibinfo {author} {\bibfnamefont {F.}~\bibnamefont
  {Wilczek}},\ }\href {\doibase 10.1103/PhysRevLett.40.279} {\bibfield
  {journal} {\bibinfo  {journal} {Phys. Rev. Lett.}\ }\textbf {\bibinfo
  {volume} {40}},\ \bibinfo {pages} {279} (\bibinfo {year} {1978})}\BibitemShut
  {NoStop}%
\bibitem [{\citenamefont {Dvali}\ \emph {et~al.}(2018)\citenamefont {Dvali},
  \citenamefont {Gomez},\ and\ \citenamefont {Zell}}]{Dvali:2018dce}%
  \BibitemOpen
  \bibfield  {author} {\bibinfo {author} {\bibfnamefont {G.}~\bibnamefont
  {Dvali}}, \bibinfo {author} {\bibfnamefont {C.}~\bibnamefont {Gomez}}, \ and\
  \bibinfo {author} {\bibfnamefont {S.}~\bibnamefont {Zell}},\ }\href@noop {}
  {\  (\bibinfo {year} {2018})},\ \Eprint {http://arxiv.org/abs/1811.03079}
  {arXiv:1811.03079 [hep-th]} \BibitemShut {NoStop}%
\bibitem [{\citenamefont {Fukuda}\ \emph {et~al.}(2019)\citenamefont {Fukuda},
  \citenamefont {Saito}, \citenamefont {Shirai},\ and\ \citenamefont
  {Yamazaki}}]{Fukuda:2018haz}%
  \BibitemOpen
  \bibfield  {author} {\bibinfo {author} {\bibfnamefont {H.}~\bibnamefont
  {Fukuda}}, \bibinfo {author} {\bibfnamefont {R.}~\bibnamefont {Saito}},
  \bibinfo {author} {\bibfnamefont {S.}~\bibnamefont {Shirai}}, \ and\ \bibinfo
  {author} {\bibfnamefont {M.}~\bibnamefont {Yamazaki}},\ }\href {\doibase
  10.1103/PhysRevD.99.083520} {\bibfield  {journal} {\bibinfo  {journal} {Phys.
  Rev.}\ }\textbf {\bibinfo {volume} {D99}},\ \bibinfo {pages} {083520}
  (\bibinfo {year} {2019})},\ \Eprint {http://arxiv.org/abs/1810.06532}
  {arXiv:1810.06532 [hep-th]} \BibitemShut {NoStop}%
\bibitem [{\citenamefont {Du}\ \emph {et~al.}(2017)\citenamefont {Du},
  \citenamefont {Behrens},\ and\ \citenamefont {Niemeyer}}]{Du:2016zcv}%
  \BibitemOpen
  \bibfield  {author} {\bibinfo {author} {\bibfnamefont {X.}~\bibnamefont
  {Du}}, \bibinfo {author} {\bibfnamefont {C.}~\bibnamefont {Behrens}}, \ and\
  \bibinfo {author} {\bibfnamefont {J.~C.}\ \bibnamefont {Niemeyer}},\ }\href
  {\doibase 10.1093/mnras/stw2724} {\bibfield  {journal} {\bibinfo  {journal}
  {Mon. Not. Roy. Astron. Soc.}\ }\textbf {\bibinfo {volume} {465}},\ \bibinfo
  {pages} {941} (\bibinfo {year} {2017})},\ \Eprint
  {http://arxiv.org/abs/1608.02575} {arXiv:1608.02575 [astro-ph.CO]}
  \BibitemShut {NoStop}%
\end{thebibliography}%
\end{document}